\documentclass{aa}

\usepackage{amssymb}
\usepackage{amsmath}
\usepackage{txfonts}
\usepackage{graphicx}
\usepackage{xspace}
\usepackage{natbib}
\usepackage{url}

\begin{document}
\title{A model for cyclotron resonance scattering features}

\author{G.~Sch{\"o}nherr\inst{1,2,3} \and J.~Wilms\inst{2} \and
  P.~Kretschmar\inst{3} \and I.~Kreykenbohm\inst{1,4}  \and
    A.~Santangelo\inst{1} \and R.E.~Rothschild\inst{5} \and W.~Coburn\inst{6} \and R.~Staubert\inst{1}}

\offprints{G.~Sch{\"o}nherr, \\ \email{gschoen@astro.uni-tuebingen.de}}

\institute{ Institut f\"ur Astronomie und Astrophysik, Abteilung Astronomie, Sand 1, 72076
 T\"ubingen, Germany \and
 Dr.~Karl Remeis-Sternwarte Bamberg, Astronomisches Institut der
 Friedrich-Alexander-Universit{\"a}t Erlangen-N{\"u}rnberg, Sternwartstr.\ 7, 96049 Bamberg, Germany \and
 European Space Astronomy Centre, ESA, Apartado 50727, 28080 Madrid, Spain \and
 INTEGRAL Science Data Centre, 16 Ch.\ d'\'Ecogia, 1290 Versoix,
 Switzerland \and
 Center for Astrophysics and Space Sciences, University of California,
 San Diego, CA 92093-0424, USA \and
 Space Sciences Laboratory, University of California, Berkeley, CA
 94720/7450, USA} 

\date{Received $<$day month year$>$; Accepted $<$day month year$>$ }

\titlerunning{A model for cyclotron resonance scattering features}
\authorrunning{G.~Sch\"onherr et al.\ }

\abstract{}{We study the physics of cyclotron line formation in the
  high-energy spectra of accreting X-ray pulsars. In particular, we
  link numerical predictions for the line profiles to results from
  observational data analysis. Therefore, first we investigate the
  theoretical predictions and the significance of our model
  parameters, and second we aim at the development of a model to
  fit cyclotron lines in observational data.}  {Simulations were
  performed using Monte Carlo methods. The data were extracted with
  HEADAS 6.1.1 and \textsl{INTEGRAL} OSA~5.1. A convolution model for the
  cyclotron line shapes was implemented for the \textsl{XSPEC} spectral
  analysis software package and for data packages compatible with
  \textsl{XSPEC} local models.}  {We predict the shapes of cyclotron lines for
  different prescribed physical settings. The calculations
  assume that the line-forming region is a low-density electron plasma, which is of cylindrical or slab geometry and
  which is exposed to a uniform, sub-critical magnetic field.  We
  investigate the dependence of the shape of the fundamental line on
  angle, geometry, optical depth and temperature. We also discuss
  variations of the line ratios for non-uniform magnetic fields. We
  have developed a new convolution and interpolation model to simulate
  line features regardless of any a priori assumed shape of the neutron
  star continuum. Fitting \textsl{RXTE} and \textsl{INTEGRAL} data of the accreting X-ray
  pulsar \object{V0332$+$53} with this model gives a qualitative
  description of the data. Strong emission wings of the fundamental cyclotron feature as
  predicted by internally irradiated plasma geometries are in principle observable by todays
  instruments but are not formed in V0332$+$53, hinting at a
  bottom illuminated slab geometry for line formation.}{}


\maketitle
\section{Introduction}\label{sect:intro}
Cyclotron Resonance Scattering Features (hereafter called CRSFs or
simply ``cyclotron lines''), first discovered in the spectrum of \object{Her X-1}
\citep{truemper77a,truemper78a}, are observed as absorption lines in the spectra of highly magnetized accreting X-ray pulsars
\citep{heindl04a}.  Their scientific importance lies in providing the
only direct method currently known for the determination of the
magnetic field of a neutron star
\citep[e.g.][]{harding06a,orlandini01a}. The line energy $E_\text{cyc}$ of
the fundamental CRSF is approximately related to the magnetic field
strength of the star by the ``$12$-$B$-$12$ rule'' 
\begin{equation}
\label{eq:12B12}
E_\text{cyc}\,\sim 11.57\,\text{keV} \cdot B_{12}\, ,
\end{equation}
where $B_{12}$ denotes the magnetic field in units of $10^{12}\,$Gauss
\citep{canuto77a}. Due to the gravitational redshift, $z$, the observed line energy
is shifted by a factor of $1/(1+z)$ with respect to $E_{\text{cyc}}$. 
Today, with the access to data from satellites as \textsl{BeppoSAX}, the Rossi X-ray Timing Explorer (\textsl{RXTE}), the International
Astrophysics Gamma-Ray Observatory (\textsl{INTEGRAL}), and \textsl{Suzaku}, the diagnostic potential of CRSFs has grown anew: with these
instruments the
observed cyclotron line shapes are energetically resolved in detail for several sources, thus demanding an
in-depth study of their formation. A better understanding of their
complex line shapes would reveal much about the physical setting
which drives the fascinating but up to today only poorly understood
processes of accretion in X-ray pulsars.
At present, the observed lines are usually modeled by Gaussian or
Lorentzian shapes \citep{mihara90a,makishima90b}. For
modeling the fundamental feature, one must sometimes use several Lorentzians or
Gaussians to obtain an acceptable fit of its non-trivial shape \citep{kreykenbohm05a}.

In this paper, using Monte Carlo simulations based on an improved
version of the code of \citet{araya99a} and \citet{araya00a}, we calculate line features for typical neutron star
spectra and infer the line
profiles under the assumption of physical parameters such as the
accretion geometry, the viewing
angle, and the plasma
temperature. Based on our simulations, we develop a new local \textsl{XSPEC}
model, \texttt{cyclomc}, for cyclotron lines and show first
results from fitting observational data with \texttt{cyclomc}.

The outline of the present work is as follows:
In Sect.~\ref{sect:overview} we summarize the basic theory of cyclotron
line formation and describe standard scenarios for the accretion process onto
the neutron star. We report key results from past observations and
give an overview of
different numerical approaches to modeling cyclotron lines for
neutron stars.
Sect.~\ref{sect:modeling} focuses on the modeling
approach taken here. We try to motivate our steps from the point of view of an
observer, however, some rather technical discussion is necessarily included
in that section.
Theoretical predictions from Sect.~\ref{sect:results} are complemented in Sect.~\ref{sect:applications} by a comparison of the model with
real data. Finally, in Sect.\ \ref{sect:summary} we summarize our
results and discuss future steps for cyclotron line modeling.

\section{Overview}\label{sect:overview}
\subsection{CRSF formation in X-ray pulsar
  spectra}\label{subsect:theory}
Cyclotron lines are found in the spectra of accreting
neutron stars in binary systems with magnetic field strengths of the order
$B \sim 10^{11}$--$10^{13}\,$Gauss. See, e.g.,
  \citet{harding06a} for a review of the physics of strongly
magnetized neutron stars. X-ray pulsars are thought to have masses of $M \sim 1.4\,M_{\sun}$, radii $R \sim 10^{6}\,$cm, and
luminosities of $10^{34}$--$10^{38}\,\text{erg\,s}^{-1}$. 
They accrete matter from a strong stellar wind or by Roche lobe
overflow, usually through an accretion disk \citep{ghosh78a}. The magnetic field of the
neutron star disrupts the flow of matter at the Alfv\'en radius, where
the magnetic field pressure equals the ram pressure of the flow, and the
matter is funneled along the field lines onto the magnetic poles
\citep{basko76a}, reaching free-fall velocities of $\sim 0.4\, c$. When discussing processes of accretion in the
following, we restrict ourselves to the field-dominated volume near
the neutron star surface at the magnetic poles.  Inverse
Comptonization of soft photons in the decelerated plasma produces photons in the X- and gamma-ray
regime. The emission characteristics of this radiation depend on the
mass accretion rate, $\dot{M}$. For large $\dot{M}$, a shock front
develops at some distance from the neutron star surface, which does
not permit the upscattered photons to escape vertically from the
accretion column, i.e., parallel to the $B$-field. As a result, a `fan beam'
emission pattern forms (Fig.~\ref{fig:geos}). As was first shown by
\citet{basko76a}, the critical luminosity for shock formation, $L^*$, is
\begin{equation}
L^{*}=2.72 \cdot 10^{37}\left(\frac{\sigma_{\text{T}}}{\sqrt{\sigma_{||}\sigma_\perp}}\right)\left(\frac{r_{0}}{R}\right)\left(\frac{M}{M_{\sun}}\right)\,\text{erg\, s}^{-1}\,,
\end{equation}
where $r_{0}$ is the polar cap radius, $\sigma_{\text{T}}$ is the
Thomson scattering cross section and $\sigma_{||}$ and $\sigma_\perp$ are the energy
averaged cross sections for the scattering of photons which
propagate in parallel and perpendicular to the magnetic field direction
\citep{becker98a}.  For small $\dot{M}$, on
the other hand, i.e., for $L<L^{*}$, the radiation is emitted from an
accretion mound such that most photons are emitted parallel to the
$B$-field in a `pencil beam' pattern.

Before emerging from the line-forming region, the high-energy photons
undergo scattering processes with the electrons in the relativistic
plasma of the accretion column. The scattering cross section is
resonant at energies determined by the separation of the Landau energies, the discrete energy levels of the
electrons: when the strength of the magnetic field $B$ approaches the critical field strength, $B_\text{crit}= (m^2c^3)/(e\hbar) = 44.14
\cdot 10^{12}\,$G, the de Broglie radius of a plasma electron becomes
comparable to its Larmor radius. Quantum mechanical treatment of the
electrons' motion perpendicular to the magnetic field lines \citep{meszaros92a,daugherty86a} reveals a
quantization of the electrons' momenta
$p_{\perp}/(m_\text{e}c)=n(B/B_\text{crit})$. This translates into
discrete energy levels, where 
the fundamental Landau level is given by the $12$-$B$-$12$
rule (\ref{eq:12B12}) and the higher harmonics have $n$
times this energy. For photon-electron scattering, relativistic
effects lead to a slightly anharmonic spacing of the resonant photon
energies. Due to the large scattering cross section
at the resonances and due to the quasi-harmonic spacing of the
thermally broadened Landau
levels, photons of energies close to the Landau level energies may not escape the line-forming region
unless inelastic scattering has slightly changed their energy. Consequently, absorption features in the photon spectrum are observed at
\begin{equation}
\label{eq:Ecycrel}
E_n = m_{\rm e} c^2 \frac{\sqrt{1+2n(B/B_{\rm crit}) \sin^2\theta}
  -1}{\sin^2\theta} \, \frac{1}{1+z} \,,
\end{equation}
where $m_{\rm e}$ is the electron rest mass, $c$ the speed of light,
$\theta$ the angle between the incident photon direction and the
magnetic field vector, and $z$ is the gravitational redshift at the
radius of the line-forming region. Note that we enumerate the
cyclotron lines starting at $n=1$, and refer to
them as the first or fundamental line at the energy $E_\text{cyc}=E_1$, followed by the second,
third, fourth, etc.\ harmonics ($n=2,3,4,\cdots$). The gravitational redshift at the
neutron star surface is approximately\footnote{Strictly spoken, Eq.~\eqref{eq:z} is exact only for a
  non-rotating, spherically symmetric, uncharged mass.}
\begin{equation}\label{eq:z}
z = \frac{1}{\sqrt{1-\frac{2GM}{Rc^2}}}-1
\end{equation}
which gives $z\sim 0.3$ assuming the typical neutron star
parameters given above. The thermal motion of the electrons parallel to the
magnetic field lines remains free and is characterized by the parallel
electron temperature, $T_\text{e}$, which will be introduced in
Sect.~\ref{subsect:physset}. 

\subsection{Observations}\label{subsect:observational}
In 1976, the first cyclotron line was detected in the
X-ray spectrum of Her X-1 \citep{truemper77a}. First interpreted as an
emission feature at $53\,$\ensuremath{\text{keV}}, the line was later
proposed to be in absorption with theoretical arguments by 
\citet{nagel81a}. Since the discovery of the Her X-1 cyclotron line, more
sources exhibiting CRSFs have been observed \citep[e.g.,][]{heindl04a,staubert03a,coburn02a,santangelo00a}. At the time of writing, more than 16 accreting
pulsars with securely detected cyclotron lines with surface magnetic fields in
the range of $1$--$5\cdot 10^{12}\,$G were known. The record holder with respect
to the number of lines detected is 4U 0115$+$63
\citep{heindl99a,santangelo99a} where five harmonics were found \citep{heindl00a}. 

CRSF sources are regular targets for observations. The progress over the last decades in energy resolution of
instruments on satellites like \textsl{BeppoSAX}, \textsl{RXTE},
\textsl{INTEGRAL}, and \textsl{Suzaku} has
led to excellent observational data of many interesting objects with complex cyclotron line features and has made high-quality phase
resolved spectroscopy possible. As a result, many interesting
characteristics of CRSFs are known today, awaiting a deeper explanation
than given by the simple picture of line formation which was outlined
in the previous section. Some key results from observational studies
are:
\begin{enumerate}
\item The profile of the fundamental line is resolved, is clearly non-Gaussian, and
  exhibits a complex shape.
\item The second harmonic generally appears deeper than the fundamental line.
\item Significant variations of the line parameters of the CRSFs with
  the pulse phase are observed for some sources.
\item The line ratios are not necessarily harmonic. The deviations
  from the harmonic energies in some spectra are too large to be
  explained only by the basic relativistic corrections implied by Eq.~\eqref{eq:Ecycrel}. 
\item The line position of the fundamental CRSF can vary with the source
  luminosity. Negative \citep{mihara95a,mowlavi06a,nakajima06a,tsygankov06a} and positive
  \citep{labarbera05a,staubert07a} linear energy-to-luminosity correlations have
  been found.
\end{enumerate}

\subsection{Numerical models}
There are two very different approaches to modeling the radiative
transfer in the accretion column: solving finite difference equations
and Monte Carlo simulations.
\subsubsection{Solving difference equations}
Motivated by the Her X-1 line detection, \citet{nagel80a,nagel81a} and later \citet{meszaros85a}
employed Feautrier methods in order to solve the radiation transfer
equation. They performed two sets of calculations, treating effects of
anisotropy and Comptonization separately. Having first presented a line-formation mechanism for a cyclotron
emission feature \citep{nagel80a}, in a later paper considering Comptonization effects,
\citet{nagel81a} then favored the Her X-1 line to appear in
absorption. For the combined effects of anisotropy and Comptonization \citet{meszaros85a} compared model predictions for different
geometries (slab and
cylinder geometry with internal or external illumination) and
discussed variations with the angle of the emergent spectra. Their
approach was later refined by the inclusion of higher
harmonics \citep{alexander91b} or by including radiation
pressure and temperature corrections in the atmosphere
\citep{bulik92a,bulik95a}. Recently, the influence of a non-uniform magnetic
field in the line-forming region on the formation of CRSFs has been
investigated with similar techniques by
Nishimura, who found a variation of the line
ratios of the CRSFs for the case of a dipolar \citep{nishimura03a} and
for a linearly varying \citep{nishimura05a} magnetic field.
\subsubsection{Monte Carlo simulations}
\citet{yahel79a} was the first to use Monte Carlo simulations for
simulating the CRSF formation in the atmosphere of a magnetized
neutron star. He considered the formation of pulse profiles and X-ray
spectra and found
that the Her X-1 feature could indeed be reproduced as a consequence of resonant
scatterings of extraordinary polarized photons. Two years later,
\citet{pravdo81a} calculated angle-dependent pulsar spectra, including
relativistic corrections to the Compton cross section and considering
polarization dependence. Focusing on the continuum spectral shape they found a
hardening of the spectra towards the magnetic equator.
\citet{wang89b} performed Monte Carlo simulations for the
geometry of a plane parallel slab with the slab normal parallel to the
$B$-field vector, and the plasma being illuminated from below. This
geometry was named `1-0 geometry' \citep{freeman99a, isenberg98b}, in contrast to the `1-1
geometry' which indicates a homogeneous slab
illuminated at the mid plane. While in the 1-0 geometry photons which
return to the source plane after scattering are absorbed (`reflected
photon flux'), in the 1-1 geometry photons may cross the source plane and
the reflected and transmitted flux are symmetric. Results from a generalized model, where the slab normal may
have any direction with respect to the $B$-field were discussed by
\citet{isenberg98a,isenberg98b}. For the case where the
slab normal is perpendicular to the magnetic field vector, their results are
comparable to assuming a cylinder geometry for the line-forming region.
\citet{isenberg98b} distinguished between line shapes of optically thin and optically thick
matter. As one key result these authors found that the line wings disappear
either for the 1-1 geometry and optically thick media or for the 1-0 geometry
and optically thin media. However, none of these scenarios could
explain the observed fundamental shallow and broad features due to the
high equivalent width of the fundamental in both cases.

Inspired by the detection of up to two cyclotron lines\footnote{Recently it was shown
  that both lines claimed by \citet{kendziorra94a} are present in the
  source spectrum
  \citep{kretschmar05a,wilson05a,inoue05a,caballero07a}.} at $50$ and
  $100\,$keV \citep{kendziorra94a} and at $110\,$keV \citep{grove95a} during subsequent outbursts of the
transient source \object{A0535+26} in 1989 and
1994, Araya and Harding presented a
new set of Monte Carlo simulations for very hard spectra of X-ray
pulsars with near-critical fields \citep{araya96a,araya99a,araya00a}. For a low-density plasma and hence low continuum optical depths,
they produced spectra for slab (1-1) and cylinder geometry of a plasma
threaded by near-critical magnetic fields and discussed the influence of
parameters as geometry, optical depth and anisotropy of the photon
source on the line shapes. The results presented in this paper are
based on their approach.

\section{Modeling CRSFs}\label{sect:modeling}
\subsection{Aims}
The key objective of this work is to obtain a physically motivated model for CRSF formation which is directly comparable to observational data. Firstly, such a comparison is fundamental when testing and reconsidering the validity of the model. Secondly, once the model has reached a well-developed state, only its simple applicability to real observational data provides the means for a systematic investigation of CRSF sources.

\subsection{Methods}
In order to achieve the desired flexibility, we base our model on
Monte Carlo simulations using a revised, generalized version of the
Araya \& Harding code. A new key feature is a Green's functions
approach giving independence from any continuum model
assumed. Araya \& Harding's implementation of an internally irradiated slab geometry is generalized to
include also the case of illumination from the bottom (1-0 geometry). More information on the Monte Carlo implementation and
details of the Green's function approach are given in Sect.\
\ref{subsect:technical}. Due to the
variety of sources and
the uncertainty of the general physical picture, calculations
are performed on a large multidimensional parameter grid. 
All simulation results are merged into archives in the form of FITS
tables which are available from the authors. Line features for X-ray pulsar spectra for different physical settings (as outlined in the
next section) within our parameter scope may be produced from these tables with a special convolution
and interpolation model, also implemented as a local model for \textsl{XSPEC} \citep{arnaud96a} and other analysis packages
such as ISIS \citep{houck00a}.

\subsection{Physical setting}\label{subsect:physset}
\begin{figure}
\begin{centering}
\resizebox{0.77\hsize}{!}{\includegraphics{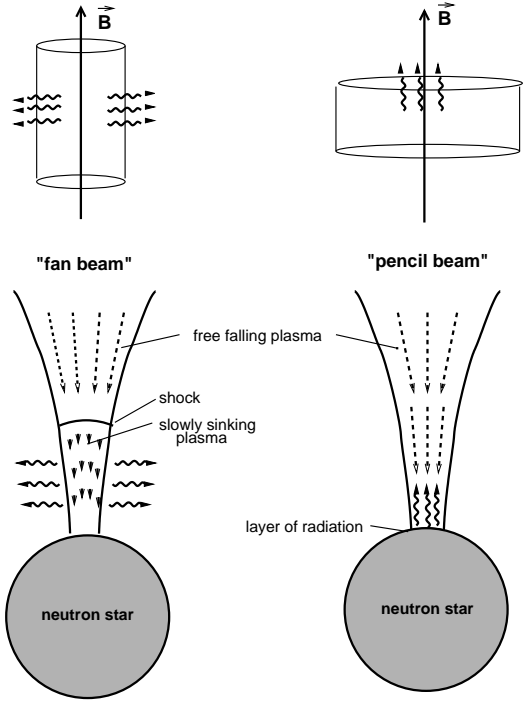}}
\caption{Accretion geometries and radiation patterns. Left: ``fan
  beam''/cylinder geometry. Right: ``pencil beam''/slab geometry.}
\label{fig:geos}
\end{centering}
\end{figure}

\begin{figure}
  \resizebox{\hsize}{!}{\includegraphics{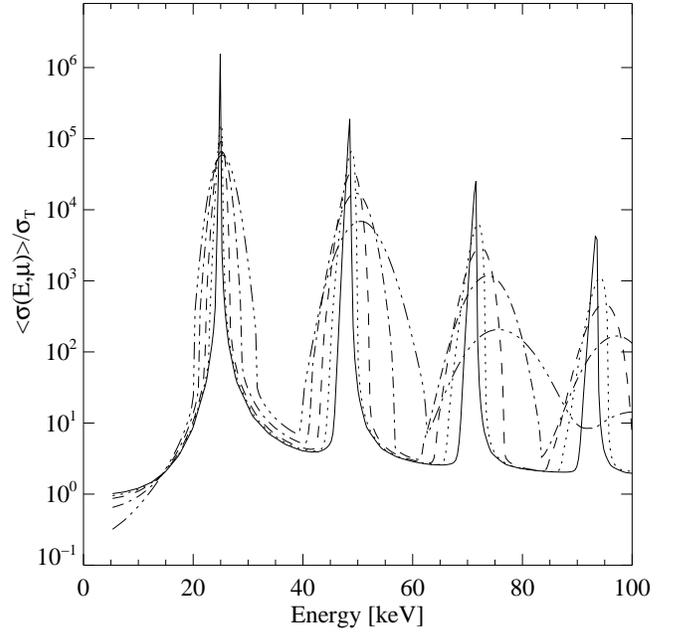}}
  \caption{Cross sections for $B/B_\text{crit}=0.05$ and $kT_\text{e}=3\,$keV,
    thermally averaged over the sampled electron momenta, and in units
    of the Thompson cross section $\sigma_\text{T}$. The resulting
    profiles are shown for different angles $\theta$ of the photon's
    direction with respect to the magnetic field vector. Solid lines:
    $\cos\theta=0.005$ and $\cos\theta=0.875$. Dotted, dashed-dotted and dashed lines
    (sharper towards broader):
    $\cos\theta=0.200,0.375,0.625$. Figure based on
    \citet{araya99a}. }
   \label{fig:profiles}
\end{figure}

We simulate the propagation of photons through a medium of prescribed
physical conditions. The photons interact with the electrons in the
medium via resonant scattering processes. The conditions
in the line-forming region of the electron plasma are governed by the
following parameters \citep{araya99a}:

\begin{enumerate}
\item \textbf{Magnetic field [$B$]}
As a first approximation we consider a neutron star with a magnetic field which is
assumed to be uniform on the scale of the line-forming region. The
field strengths simulated are between $1\cdot 10^{12}$ and $7 \cdot 10^{12}$ Gauss, encompassing the whole
range of $B$-fields found in cyclotron line 
sources. In Sect.~\ref{subsect:res:B} a possible generalization to
non-uniform magnetic fields is described.
\item \textbf{Plasma electrons [$T_\text{e}, f(p_\text{e}),n$]}
We consider a low-density thermal plasma. We also assume that all
electrons are initially in their fundamental Landau state $n=0$. This
assumption is justified by the very high cyclotron radiative decay rate for
sub-critical fields 
\begin{equation}
r_{\text{rad}}=3\cdot10^{15}\,B_{12}^{2}\,\text{s}^{-1}
\end{equation}
compared to the collisional excitation rate
\begin{equation}
r_{\text{col}}=5\cdot
10^{8}(n_\text{e}/10^{21}\text{cm}^{-3})\,B_{12}^{-3/2}\,\text{s}^{-1}
\end{equation}
\citep{latal86a,bonazzola79a}. For their motion parallel to the
$B$-field vector, we assume a thermal distribution
of the electrons with their parallel
momenta $p_\text{e}$ given by a relativistic Maxwellian distribution
\begin{equation}
\label{eq:fp}
f(p_\text{e})dp_\text{e} \propto
\exp{\left(-\frac{m_\text{e}c^2\left(\sqrt{1+\left(\frac{p_\text{e}}{m_\text{e} c}\right)^2}-1\right)}{kT_\text{e}}\right)dp_\text{e}} \,,
\end{equation}
where $T_\text{e}$ is the parallel electron temperature and $k$ is the Boltzmann constant. In
the literature, $T_\text{e}$ is often linked to the
strength of the magnetic field \citep[e.g.][]{lamb90a, isenberg98b, araya96a, araya99a,araya00a}. We take these studies into account in order to determine the order of
magnitude of the plasma temperature, but leave $T_\text{e}$ as a free
parameter in our
simulations in order to keep the model's flexibility (see also
Sect.~\ref{subsect:res:Te}). We also assume a slowly sinking plasma where
bulk plasma motion may be neglected.
\item \textbf{Optical depth [$\tau_\text{T}$]}\label{subsubsect:tau}
The Thomson optical depth $\tau_\text{T}$ of the plasma is
prescribed. The optical depth for cyclotron scattering $\tau_\text{cyc}$ relates to the Thomson optical depth as
\begin{equation}
\tau_\text{cyc} = \frac{\sigma_\text{cyc}}{\sigma_\text{T}}\tau_\text{T}\,,
\end{equation}
implying a scattering optical depth $\tau_\text{cyc}$ which can be a factor of $\sim$$10^{5}$ larger than $\tau_\text{T}$ at the resonances of
$\sigma_\mathrm{cyc}$. Fig.~\ref{fig:profiles} shows the thermally
averaged cross section $\langle \sigma_\text{cyc}\rangle/\sigma_\text{T}$ as a function of
energy and angle, calculated as a second order QED process
\citep{sina96a}. Besides a highly resonant behavior of the cross
section at the Landau energies, Fig.~\ref{fig:profiles} also
illustrates the angle-dependent relativistic shift in the resonances
as well as the thermal broadening of the profiles. We calculate line
features for Thomson optical depths between $\tau_\text{T}=1\cdot 10^{-4}$
and $\tau_\text{T}=3\cdot 10^{-3}$ \citep{araya99a}. Depending on the plasma geometry, the
resulting mean free path of a photon in the line-forming region is
different for the same trajectory.  The Thomson optical depths we
simulate correspond to electron column densities
$N_\text{e}=\tau_\text{T}/\sigma_\text{T}$ between $1.5\cdot 10^{20}\,
\text{cm}^{-2}$ and $4.5\cdot 10^{21}\,\text{cm}^{-2}$. These values are comparable to values assumed in other recent 
numerical or analytical studies. For instance, column densities of 
$N_\text{e}\sim 10^{21}$--$10^{22}\, \text{cm}^{-2}$ are assumed by
\citet{nishimura03a,nishimura05a}, and values of $N_\text{e}\sim 10^{22}\, \text{cm}^{-2}$ are inferred by
\citet{becker07a} for the sources Her X-1, \object{LMC X-4}, and \object{Cen X-3}.

\item \textbf{Geometry [geo: sl/cy] \label{subsubsect:geo}} We
  distinguish two basic geometries of the line-forming region motivated by the complementary `standard' pictures of accretion depicted in
  Fig.~\ref{fig:geos} \citep[see also section \ref{subsect:theory}
  and][]{basko76a}: for the case of flow stopping through nuclear
  collisions at the surface, we adopt the geometry of a thin,
  plane-parallel slab \citep{meszaros83a,harding84a}. Radiative shocks or shocks from
  collisionless instabilities, on the other hand, require a cylindrical
  shape of the X-ray emitting region \citep[e.g.,][and references
  therein]{becker07a}. The heights of slab atmospheres are expected to
  be significantly smaller \citep{lamb73a,wang88a} than typical radii,
  $h \ll r_0$, of the
  accretion mound \citep{ostriker73a,becker07a}. For
  cylindrical geometries, \citet{becker07a} investigated a more
  complicated radiative shock structure with a
  velocity gradient. For several X-ray pulsars, these authors found height to width ratios, $z_\text{max}/r_0$, of the emitting
  region of the order $10^2$, confined to a
  volume smaller than the one of trapped radiation. Our
  simulations also suggest that practically all photons escape before
  reaching such boundaries, justifying our assumption of infinite
  extended plane-parallel line-forming regions. For the cylinder, we assume that the
  plasma is irradiated internally with the photon source located at
  the cylinder axis \citep{araya99a,araya00a}. Two locations of the
  source plane are considered for the slab. The scenario of a line formation region above an isotropically
  emitting source is realized by a bottom-illuminated slab
  \citep{freeman99a,isenberg98b,wang89a}, while placing the photon source at the midplane of a
  plane-parallel slab is representative of the
  scenario of line formation in an isothermal, semi-infinite
  atmosphere \citep{slater82a}.

The optical depth the photons see into direction $\theta$
  when covering an optical depth $\Delta\tau_\text{T}$ along the slab normal or
  perpendicular to the cylinder depends on the geometry as:
\begin{align}
\Delta\tau_\text{T}(\theta) &= \frac{\Delta\tau_\text{T}}{\sin\theta} \quad \text{(cylinder
  geometry)} \label{eq:taucy}\\
\Delta\tau_\text{T}(\theta) &= \frac{\Delta\tau_\text{T}}{\cos\theta} \quad \text{(slab
  geometry)\, .} \label{eq:tausl}
\end{align}  

\end{enumerate}

\subsection{Technical realization}\label{subsect:technical}
We use a Monte Carlo method to simulate the resonant scattering
processes between incident photons and plasma electrons which lead to
the formation of cyclotron line features.  The code used is a revised
version of the code of Araya \& Harding \citep{araya96a,araya99a,araya00a}. The calculation of the
relativistic cross sections is done with a separate code by \cite{sina96a}. Resonant scattering with electrons up to the fourth harmonic is
included.

Besides relaxed geometrical constraints on photon injection, and technical modifications such as an improved angular and energy
resolution and increased statistics, the main difference of our
program with respect to the preceeding one is the Green's functions
approach: In each Monte Carlo run we insert $10000$ photons of the
same incident energy, $E_\text{in}$, pick for each photon a random angle $\theta_\text{in}$
($\cos(\theta_\text{in}) \in (-1,1))$ with respect to the magnetic
field direction, propagate them through the plasma
(see below), and calculate from all final
states the probabilities for photon redistribution into different
energy and angular bins. The initial angular distribution of the
photons is assumed to be isotropic. The relevant energy range
$\left\{E_\text{in}\right\}$ for cyclotron line formation is assessed as follows: using the $12$-$B$-$12$ rule and assuming quasi-harmonic spacing
of the cyclotron lines, the energy range containing the first four
Landau levels can be fixed independently from the magnetic field
strength in terms of $E/B_{12}$. Due to the link of resonant energy and
magnetic field, the choice of this scale is important. First of all, it gives an optimized resolution of the CRSF features in the same way for
different magnetic fields. Even more important, the choice of this
scale is fundamental for later interpolation of the Green's
functions: as the resonant energies are directly linked to the magnetic
field, an interpolation of the line shapes in $E/B_{12}$-space
ensures consistent results. For
$E_{\text{in}}/B_{12}/\text{keV} \in [6,48]$, we obtain a grid of
Green's functions $G(E_{\text{in}}\rightarrow
E_\text{out},\theta_\text{out})$. $E_\text{in}$ is sampled by 161 Monte
Carlo runs, the resolution of the redistributed energies $E_\text{out}$ is
given by an internal energy binning of 640 bins.

Each Monte Carlo photon is injected into the plasma with energy
$E_{\text{in}}$ and direction $\theta_\text{in}$. The photon is then propagated according
to its mean free path $1/(n_\text{e}\,\langle\sigma(E_\text{in},\cos\theta_\text{in})\rangle\,)$ and an
electron is picked as a scattering partner. The electron is characterized by its
parallel momentum, $p_\text{e}$ (Eq.~\ref{eq:fp}), and its Landau
state $n$. According to the scattering cross section (obtained from interpolation of
previously calculated and tabulated values as a function of $B$), the
state of the electron-photon pair changes from its incident
configuration $(E^{(0)},\theta^{(0)})+(p_\text{e}^{(0)},n^{(0)})$ to a
different state $(E^{(1)},\theta^{(1)})+(p_\text{e}^{(1)},n^{(1)})$. The new mean free path of
the photon is calculated and the photon
is propagated further. If the electron remains in an excited Landau state
$n' > 0$ after scattering, another photon is
emitted with $(E^{(2)},\theta^{(2)})+(p_\text{e}^{(2)},n^{(2)})$ and processed further. This photon spawning can produce up to three
secondary photons.  Once a photon has escaped from the plasma, its contribution to the output spectrum is stored. Fixing the input angular distribution of the incident
photons to be isotropic in the scope of the present work is done for
reasons of simplicity, and in order to keep the computational expenses
reasonable. In a future paper we will discuss the generalization to
arbitrary angular photon distributions. Polarization of the photons is
not included, however, polarized photons should yield a comparable
picture \citep{wang88a} for the low-density regime chosen.

\begin{figure}
  \resizebox{\hsize}{!}{\includegraphics{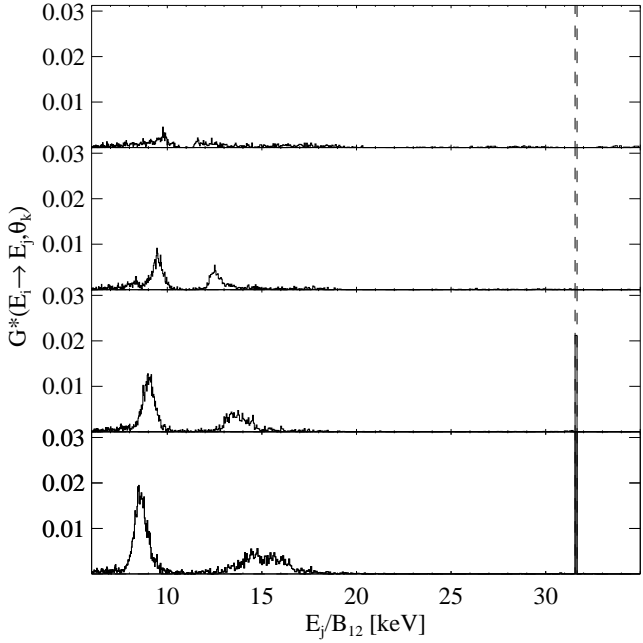}}
  \caption{Green's functions plotted for one input energy
    $E_i/B_{12}/\text{keV}=31.6$ (dashed lines) against all output energies $\{E_j\}_j$, and into four (of eight) $\cos\theta$ bins
    ($\cos\theta \in
    [0.125,0.250)$, $[0.375,0.500)$, $[0.625,0.750)$, $[0.875,1.0)$
    from top to bottom). The
    physical setting is as follows: geo:sl, $B/B_\text{crit}=0.06$,
    $kT_\text{e}=5\,$keV, $\tau_\text{T}=3\cdot 10^{-3}$. $E_i$
    corresponds to photon input at the third harmonic line. Most
    photons are redistributed by photon spawning to the wings of the
    fundamental line.}
   \label{fig:greens_zoom_lin}
\end{figure}

Calculations were performed in six-dimensional parameter space for a non-regular grid of points
$[B/B_\text{crit},T_\text{e},\mu,\tau_\text{T},E_\text{in},\text{geo}]$ within
the ranges $B/B_\text{crit}\in [0.03,0.15]$, $kT_\text{e} \in [2.5,20]\,$\ensuremath{\text{keV}},
$\mu=\cos(\theta_\text{out}) \in (0,1)$, $\tau_\text{T} \in
[1\cdot10^{-4},3\cdot 10^{-3}]$, $E_\text{in}/B_{12}/\text{keV} \in [6,48]$),
and for slab 1-1, slab 1-0, and
cylinder geometry respectively. The current parameter grid is resolved into
$(N_{B/B_{\text{crit}}}\times N_{T_{\text{e}}}\times N_{\mu}\times
N_{\tau_{\text{T}}}\times N_{E_{\text{in}}}\times
N_\text{geo})=(16\times 4\times 8\times 4\times 161\times 3)$ grid
points requiring a simulation time of the order of
  $10^{5}$ CPU hours on $2\,$GHz workstations.
The resolution was chosen such that the variation of the Green's
functions between two points is sufficiently small to allow for
interpolation. Hence, we can predict CRSFs 
by convolution of a continuum spectrum for any parameter combination
($B/B_\text{crit},T_\text{e},\mu,\tau_\text{T},\text{geo}$)
on this grid as follows:
First, we obtain the corresponding Green's functions G$^*$ by linear
interpolation in all parameters except the geometry on
$E/B_{12}$. A set of Green's functions $\{G^*(E_i \rightarrow E_j,\theta_k)\}_j$ for a fixed example
input energy $E_i$ and fixed physical setting is shown in
Fig.~\ref{fig:greens_zoom_lin}. Second, we calculate the emergent photon flux $F^\text{em}(E_j,\theta_k)$, i.e.,
the number of photons per \ensuremath{\text{keV}} in the $j^\text{th}$
energy bin and $k^\text{th}$ angular bin (binned in $\cos\theta$), as
a function of the (isotropic) incident continuum flux $F^\text{cont}$ by
convolving $F_\mathrm{cont}$ with this interpolated set of Green's
functions
\begin{equation}
F^\text{em}(E_j,\theta_k)=\frac{\sum_i G^*(E_i \rightarrow E_j,\theta_k)\, F^\text{cont}(E_i)\Delta E_i}{\Delta E_j}\,.
\end{equation}
Note that because of this approach CRSFs for arbitrary continuum 
shapes can be calculated without rerunning the simulations.

\subsection{X-ray pulsar continua}
In the following, we show cyclotron resonance scattering features in folded full spectra. As continuum
input, we chose an exponentially cutoff power law of the form
\begin{equation}
F(E) = A \cdot E^{-\alpha}\exp{(-E/E_\text{fold})}
\end{equation}
where $A$ is the power law normalization, $\alpha$ the photon index
and $E_\text{fold}$ the folding energy. This spectral shape is the
most simple phenomenological model which qualitatively describes X-ray
pulsar spectra and is used here for illustrative purposes only. When
considering real observational data, however, more complex continuum
models like, e.g.\, a power law with a Fermi-Dirac cutoff
\citep{tanaka86a} or a negative and positive powerlaw with exponential
cutoff (`NPEX') \citep{mihara95a} can be used. 
Only recently, an analytical derivation of the
spectral shapes of X-ray pulsar spectra was presented by
\citet{becker05a, becker07a}.

\section{Results}\label{sect:results}
In this section we make theoretical predictions from our Monte
Carlo simulations and discuss their implications on observed
properties of cyclotron lines. Special emphasis is placed on the study of the
line parameters, i.e., line position, line width and line depth, of the fundamental CRSF.

\begin{figure}
  \resizebox{\hsize}{!}{\includegraphics{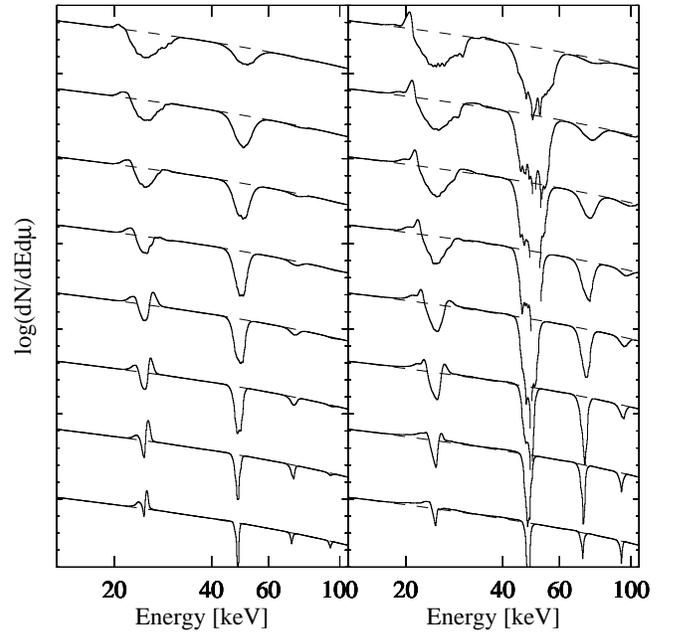}}
  \caption{Line profiles for cylinder geometry as a function of angle
    and optical depth. Left: continuum
    optical depth $\tau_\text{T} = 3 \cdot 10^{-4}$; right: $\tau_\text{T} = 3
    \cdot 10^{-3}$. The emergent spectra [photons ster$^{-1}$
    s$^{-1}$ keV$^{-1}$] from all eight angular bins are shown (bottom
    to top: $\mu = \cos\theta \in [0.000,0.125)$, $[0.125,0.250)$,
    $[0.250,0.375)$, $[0.375,0.500)$, $[0.500,0.625)$,
    $[0.625,0.750)$, $[0.750,0.875)$, $[0.875,0.1000)$). In both
    panels, $B/B_\text{crit}=0.05$ and $kT_\text{e}=3.0$\,keV. The
    continuum photon flux is assumed to have a power law distribution
    with photon index $\alpha= 2.0$ and an exponential high energy
    cutoff at the folding energy $E_\text{fold}=40\,$\ensuremath{\text{keV}}.}
   \label{fig:fullspectra_cy_tau}
\end{figure}

\begin{figure}
  \resizebox{\hsize}{!}{\includegraphics{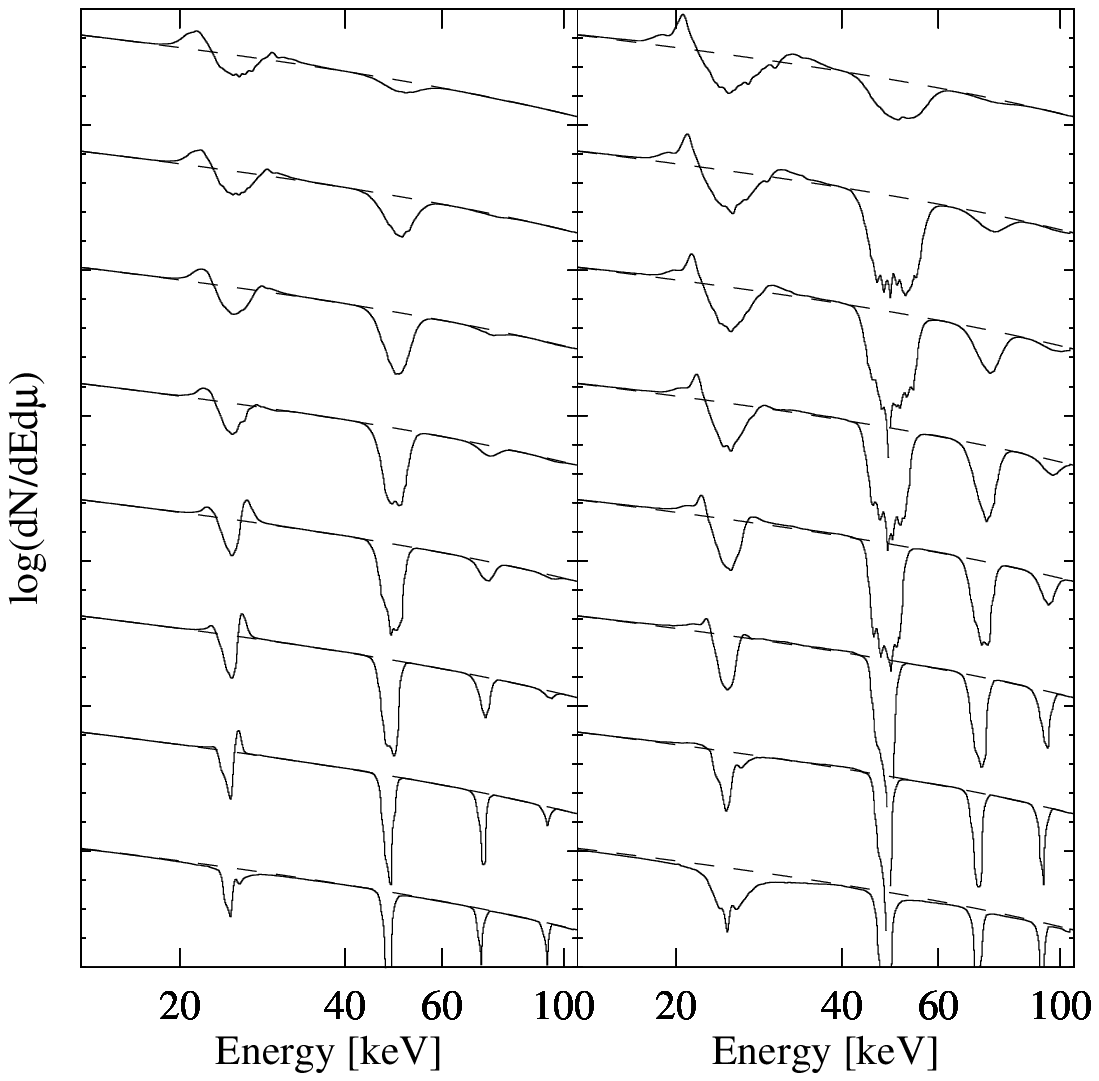}}
  \caption{Same as Fig.\ \ref{fig:fullspectra_cy_tau} for slab 1-1 geometry.}
   \label{fig:fullspectra_sl_tau}
\end{figure}

\begin{figure}
  \resizebox{\hsize}{!}{\includegraphics{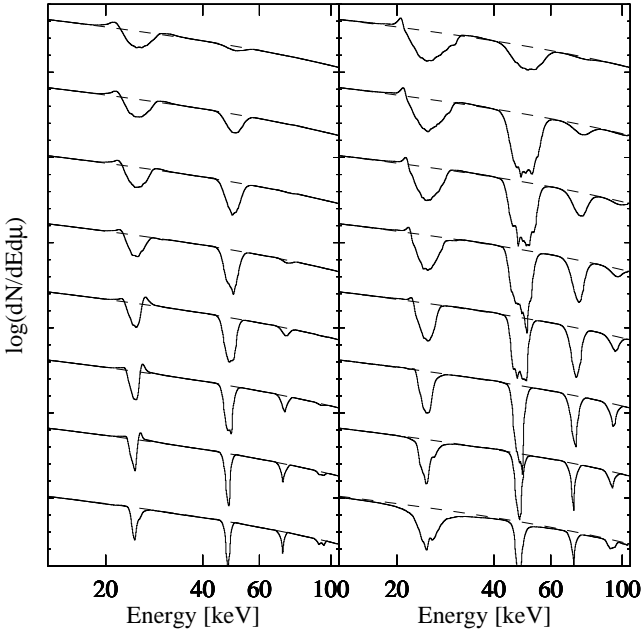}}
  \caption{Same as Fig.\ \ref{fig:fullspectra_cy_tau} for slab
      1-0 geometry.}
   \label{fig:fullspectra_sl10_tau}
\end{figure}

\subsection{Geometry and optical depth}
\label{subsect:res:geo}
To close the link to the publications by \citet{araya99a} and \citet{araya00a}, we
illustrate in Figs.~\ref{fig:fullspectra_cy_tau}
and~\ref{fig:fullspectra_sl_tau} full spectra folded with the
convolution model for a chosen physical setting in a similar fashion
as shown in their work: spectra are depicted for slab and cylinder
geometries, for two different values of $\tau_\text{T}$ and for different
angular bins. The continuum parameters are fixed to
$E_\text{fold}=40\,$keV and $\alpha=2$. The choice of the folding
energy describes a rather hard continuum, which leads to
a pronounced contribution from high-energy photon spawning to the
CRSFs (compare Sect.~\ref{subsect:cont}). Our results are in agreement
with the previous simulations of \citet{araya99a}. As an improvement with respect to
the earlier results, we calculated the line shapes on a grid of 640
energy bins instead of 80 and we resolve the depicted line shapes for
eight angular bins of $\cos\theta$ instead of four. The improved
statistics of effectively $1.6\cdot 10^{6}$ Monte Carlo photons per folded spectrum
(increased from $5\cdot 10^4$) gives us a well resolved picture of
the line shapes, however, the centroid of the second harmonic would
require even better statistics for continuum optical depths of
$\tau_\text{T} \sim 3\cdot 10^{-3}$. Although the second harmonic is not
fully described in its core, the calculations clearly show that the
second harmonic is more pronounced than the fundamental one, which has a
more complex, broad and shallow shape, often with emission wings,
  which are strongest for the internally illuminated plasmas. The strength
of those
emission wings is expected as a consequence of the source photon
  location at the mid-plane of a slab and
  mid-axis of a cylinder. For slab geometry, \citet{isenberg98b} and
\citet{nishimura05a} showed comparisons of the slab 1-1 and
the slab 1-0 geometry, and pointed out that
these wings can be understood as the remnants of a strong emission
feature forming near the mid-plane. For very small optical depths, $\tau_\text{T}=10^{-4}$, we also find this feature in our simulations.
Observations of sources with CRSFs have not been seen to exhibit strong emission wings. In Sect.~\ref{subsect:observability}, we will
comment further on the observability of those features with modern
instrumentation.

\begin{figure}
  \resizebox{\hsize}{!}{\includegraphics{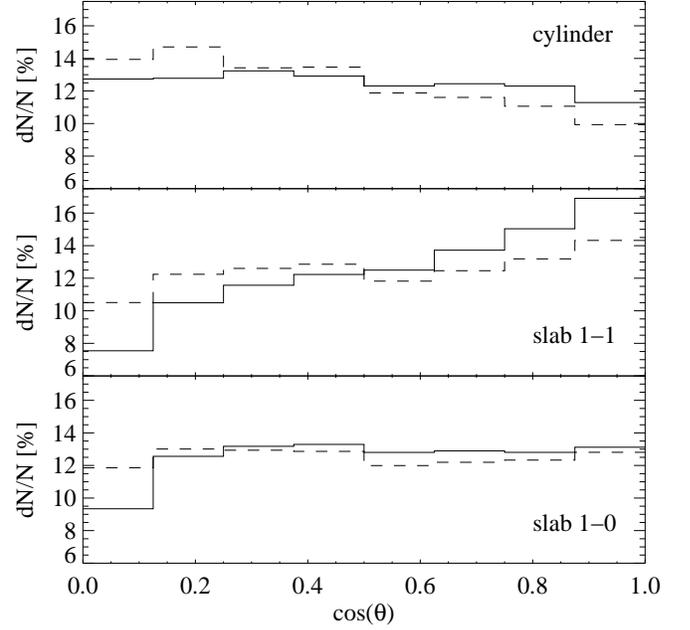}}
  \caption{Angular redistribution of the photons. For isotropic photon
    injection, the percentage of the emitted photon flux into eight
    angular bins for the spectra in Figs.\
    \ref{fig:fullspectra_cy_tau}, \ref{fig:fullspectra_sl_tau} and \ref{fig:fullspectra_sl10_tau} is shown. Top:
    cylinder geometry, middle: slab 1-1 geometry, bottom: slab 1-0 geometry. Solid lines:
    $\tau_\text{T}=3\cdot10^{-3}$, dashed lines: $\tau_\text{T} = 3\cdot 10^{-4}$. }
   \label{fig:ang_redist}
\end{figure}
\begin{figure}
  \resizebox{\hsize}{!}{\includegraphics{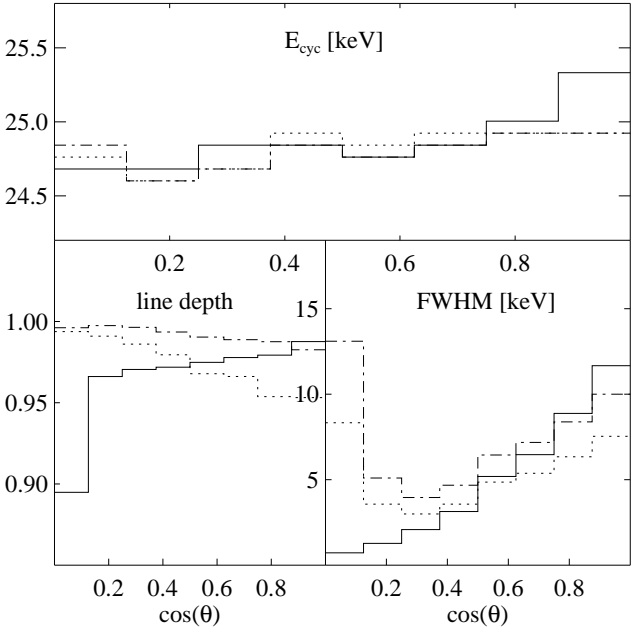}}
  \caption{Variation of the line parameters of the fundamental CRSF with
    the angle of emission. Solid lines: cylinder geometry, dashed lines:
    slab 1-1 geometry, dash-dotted lines: slab 1-0 geometry. All values are obtained from fits of the line
    shapes depicted in Figs.\
    \ref{fig:fullspectra_cy_tau}, \ref{fig:fullspectra_sl_tau} and \ref{fig:fullspectra_sl10_tau}, and for
    a Thompson optical depth of $\tau_\text{T}=3\cdot 10^{-3}$.}
   \label{fig:ang_linevary}
\end{figure}

\subsection{Angular distribution}
\label{subsect:res:mu}
Photons are injected isotropically into 20 angular bins in
$\cos\theta$. Although the distribution of the initial photon directions is
isotropic, a high degree of anisotropy arises after the photons
have been propagated through the plasma due to a highly anisotropic
scattering cross section. Fig.~\ref{fig:ang_redist} shows the
angular redistribution of the photons for all considered geometries and for
different values of the optical depth. Internally the code keeps track
of twenty $\cos\theta$ bins. We show the angular redistribution
resolved into
eight final bins. For $\tau_\text{T}=3\cdot 10^{-4}$ we observe a trend of an overall
redistribution towards smaller $\theta$ for slab geometry and a
reverse trend for cylinder geometry. For a larger optical depth, $\tau_\text{T}=3\cdot 10^{-3}$, these trends increase for slab and decrease
for cylinder geometry, where the curve in
Fig.~\ref{fig:ang_redist} flattens. This can be understood from
the dependence of the scattering cross sections on the angle (see
Fig.~\ref{fig:profiles}) which implies that there is general trend of
a photon redistribution by scattering towards smaller angles, i.e., larger
$\cos\theta$, regardless of the geometry. The larger the optical depth a
photon must pass, the more scatterings take place
and the more dominant this effect becomes. This condition can also account for a
  less prominent trend in the slab 1-0 geometry compared to the 1-1 geometry,
  as photons which experience many scatters by various crossings of the source
  plane are thermalized, biasing the emerging radiation.
For a fixed optical depth $\tau_\text{T}=1\cdot 10^{-3}$ similar plots are shown resolved into all
twenty bins by \citet{araya00a}. For $\tau_\text{T}\sim 8\cdot 10^{-4}$, the angular photon redistribution of the
line photons is discussed by \citet{isenberg98b}.
Due to the redistribution of photons in angle, the cyclotron
line shapes also vary significantly. We have fitted the fundamental feature
with the continuum spectral function multiplied with a Lorentzian in absorption for the
absorption feature and two Lorentzians in emission for the emission
wings. Fig.~\ref{fig:ang_linevary} shows characteristic
parameters of the fundamental CRSF as obtained
from these phenomenological fits.

The line position of the fundamental CRSF varies little with
the viewing angle and thus cannot account for the amount of change in
line positions observed for some sources with phase or during
various observations. Therefore, different explanations have to be
sought. A possible scenario could be a variation of the magnetic field with
angle. The line depth increases with $\cos\theta$ in the case of cylinder
geometry, and decreases for slab geometry. This is easily understood
from Eqs.~\eqref{eq:taucy} and \eqref{eq:tausl}, as they predict the
largest optical depth for small $\theta$ for cylinder, and for
large $\theta$ for slab geometry. For cylinder geometry, the line width
increases clearly with $\cos\theta$ as expected from the angle-dependence of
relativistic Doppler broadening (see Eq.~\ref{eq:FWHM} below) and
reinforced by the increasing optical depth with $\cos\theta$. The decreasing optical depth with
angle for slab geometry instead suppresses the trend of the line broadening
with $\cos\theta$.

\begin{figure}
  \resizebox{\hsize}{!}{\includegraphics{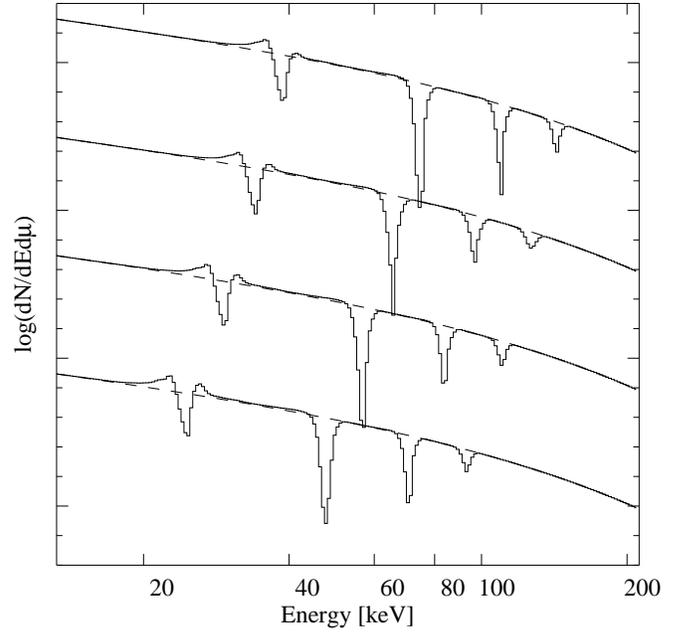}}
  \caption{Line profiles for different values of the magnetic field strength. The $B$-field
    increases from bottom to top:
    $B/B_\text{crit}=0.05,0.06,0.07,0.08$. We assume cylinder
    geometry, a constant temperature
    $kT_\text{e}=3\,$\ensuremath{\text{keV}}, an optical depth
    $\tau_\text{T}=3\cdot10^{-3}$ and a viewing angle $\theta$ within $\mu=\cos\theta=0.25$. The continuum
    spectrum has the shape of a power law with photon index $\alpha = 2.0$
    and with an exponential cutoff at the energy $E_\text{fold}=40$ keV.}
   \label{fig:fullspectra_cy_B}
\end{figure}
\begin{figure}
  \resizebox{\hsize}{!}{\includegraphics{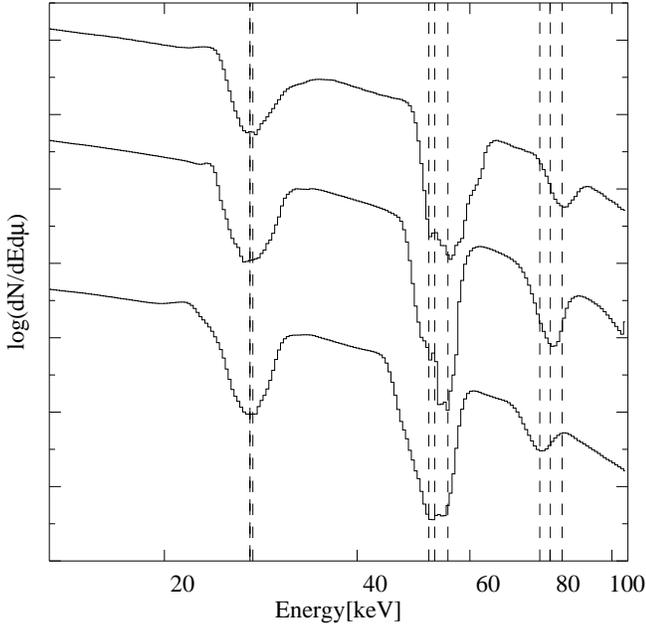}
  }
  \caption{Comparison of CRSFs for a uniform and
  a non-uniform magnetic field in a line-forming region of
  slab 1-0 geometry. Bottom:
  $B/B_\text{crit}=0.05\rightarrow 0.055$, Middle: $B/B_\text{crit}=\text{const.}=0.055$, Top:
  $B/B_\text{crit}=0.06\rightarrow 0.055$. Results are shown for
  $\mu=\cos\theta=0.6$. Otherwise the same setting as in Fig.\
  \ref{fig:fullspectra_cy_B} is used. The vertical dashed lines mark the line
  positions obtained from a phenomenological fit (continuum multiplied
  by three Lorentzians in absorption for the lines and two Lorentzians
  in emission for the emission wings of the fundamental line) of the first three CRSFs.}
   \label{fig:Bvarying}
\end{figure}

\subsection{Line energies vs.\ magnetic field strength $B$}
\label{subsect:res:B}
Line profiles for different magnetic field strengths and otherwise
fixed parameters are shown in Fig.~\ref{fig:fullspectra_cy_B}. The figure clearly shows the approximately linear
progression of the centroid line energies towards higher energies with
increasing magnetic field strength expected from the 12-$B$-12 rule.
The relativistically correct line ratios for a uniform field are implied by
Eq.~\eqref{eq:Ecycrel}. For some sources, e.g., V0332$+$53 \citep{pottschmidt05a} this behavior is confirmed by
observational data.
As mentioned above, there are also sources for which the line energies of
the harmonics are much less harmonically spaced than can be
accounted for by relativistic effects. One possible explanation is a
model in which the magnetic field is varying locally within the
line-forming region. Estimates of the dipole
  magnetic field of pulsars from torque theory and measurements of the surface magnetic
  field from CRSF detections hint at a magnetic field of more complex structure
  than a simple dipole. As the exact nature of the magnetic field is
  unknown, different scenarios for non-dipole magnetic field structure have been
  investigated \citep{blandford83a,urpin86a,arons93a}. Besides magnetic field gradients which arise due to
  dipole variations with the height of the magnetosphere, surface field
  variations can result, e.g., from small-scale crustal field structures
  \citep{blandford83a}, or thermomagnetic field evolution effects \citep{urpin86a}. \citet{nishimura05a} recently presented a
  study of the line ratios for a line-forming region of slab geometry
  threaded by a magnetic field which linearly varies with the height. Their approach was based on a model by \citet{gil02a} who assumed the
  presence of a star-centered dipole from a fossil field in the core
  superposed by a crust-anchored dipole anomaly from crustal field
  structures. Applying Feautrier methods to solve the radiative transfer, \citet{nishimura05a} found that the line ratios
significantly increase if the $B$-field decreases upwards, and
decrease vice versa. We have adapted this toy model for a first study
of magnetic field spread with our Monte Carlo approach. With some simplifying assumptions considering the
geometry and the angular redistribution for the case of a non-constant
$B$-field, we can confirm the trend of line-ratio increase and
decrease proposed by \citet{nishimura05a}. An example is shown in
Fig.~\ref{fig:Bvarying}, where we compare cyclotron lines for a
constant, a linearly decreasing, and a linearly increasing magnetic
field. For the non-constant magnetic field we prescribe a linear
variation of the field strength in discrete steps within the line
forming region of up to 10\%.  The fundamental line appears widely
unchanged in shape and position, as it is formed in the upper scattering
layers, where the non-constant $B$-field was set to have the same
value as the constant one. Line photons which have been scattered out of the line of
sight or redistributed in energy in lower layers are replaced
by spawned photons from scattering in higher layers. By contrast, the higher harmonic lines
change in position and shape. Here, contributions from all layers of
different depth are
important for the final line profile. Absorption features from photons
at low layers are not refilled. Hence, with the change of the resonant
energies with the height of the line forming region the lines become
wider (proportional to the amount of variation in $B$) and their final 
centroid energy is shifted. From fitting Lorentzians to the first
three lines we obtained the line energies and line ratios for a
constant, increasing and decreasing magnetic field. In
Table~\ref{tab:linepos} we list the fitted line positions and the fitted line ratios. The theoretical values of the
line energies and line ratios after Eq.~\eqref{eq:Ecycrel} are shown
for completeness. As expected, these values are similar to the fitted
ones for a constant magnetic field, although not identical, as the
non-Gaussian and non-Lorentzian line shapes especially of the
fundamental line restrict the fit quality noticeably.

\begin{table}
  \caption{Comparision of line positions and line ratios from the fits of
    the spectra shown in Fig.\ \ref{fig:Bvarying}. The line energies are
    given in keV. For the case of a constant magnetic field, the line
    energies are also calculated from Eq.\ (\ref{eq:Ecycrel}).} 
\label{tab:linepos} \centering
\begin{tabular}{c c c c c }\hline\hline
$B/B_\text{crit}$ & $1^\text{st}$ & $2^\text{nd}$ & $3^\text{rd}$ &
    line ratios \\ \hline
$0.050 \rightarrow 0.055$ & $27.20$ & $51.76$ & $77.16$ & $2.84:1.90:1$ \\ 
$0.055$ & $27.20$ & $52.87$ & $80.14$ & $2.95:1.94:1$ \\ 
$0.060 \rightarrow 0.055$ & $27.49$ & $55.45$ & $83.61$ &
$3.04:2.02:1$ \\
$0.055$ (Eq.~\ref{eq:Ecycrel}) & $27.62$ & $54.35$ & $80.27$ &  $2.91:1.96:1$ \\
\hline
\end{tabular}
\end{table}
Line ratios of strongly anharmonic nature have been observed. For \object{4U 0115$+$63}, the observed spacing of the
line energies is smaller than expected, yielding line ratios of
($2.8\pm 0.05:1.9\pm 0.05:1$) \citep{santangelo99a} or ($2.71\pm
0.13:1.73\pm 0.08:1$) \citep{heindl99a} for the first three harmonics.
These deviations are comparable to the ones we obtained for an
increasing magnetic field in our example. Using a different modeling approach, and tuning
his parameters to the case of 4U 0115$+$63, \citet{nishimura05a} has already
shown that these line ratios can be reproduced by the assumption of a
variable magnetic field in the line forming
region.
An example for a source where the line ratios may be higher than what is
expected from Eq.~\eqref{eq:Ecycrel} is \object{Vela X-1}, where \citet{kreykenbohm99a,kreykenbohm02a} have
found a coupling of the first harmonic energy to the fundamental line
energy $\gtrsim 2$ in \textsl{RXTE} data.

\subsection{Influence of the plasma temperature}\label{subsect:res:Te}
\begin{figure}
  \resizebox{\hsize}{!}{\includegraphics{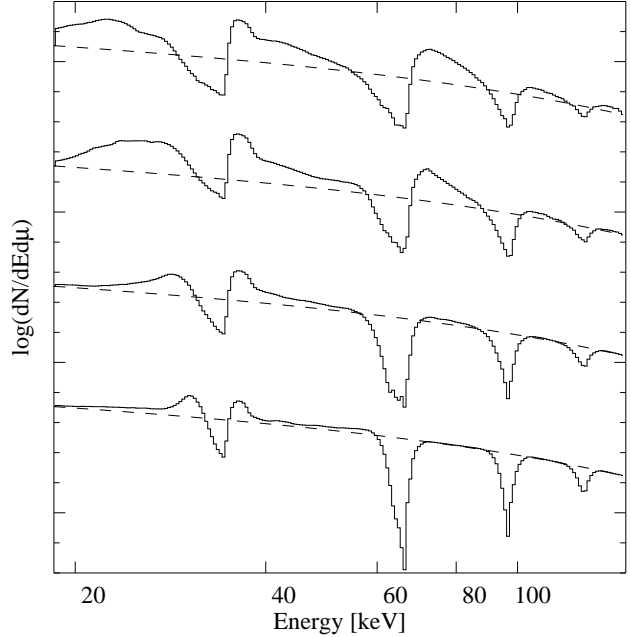}
  }
\caption{Variation of the line shapes with the temperature
  ($kT_\text{e}=5,10,15,20\,$keV from bottom to top). Spectra are shown for
  a magnetic field $B/B_\text{crit}=0.07$ and otherwise the same
  parameters as in Fig.~\ref{fig:fullspectra_cy_B}.} 
   \label{fig:fullspectra_cy_Te}
\end{figure}
The line width is determined by the energy and angle-dependent shape
of the scattering cross section and smeared out due to thermal Doppler
broadening.  In Fig.~\ref{fig:fullspectra_cy_Te}, cyclotron line
shapes are depicted for varying parallel electron temperature $T_\text{e}$,
for cylinder geometry, fixed parameters $B$, $\cos\theta$ and for a fixed continuum
shape.  The hotter the plasma, the wider and the more
asymmetric become the lines. The width of the lines is due to a
combination of the natural line width and Doppler
broadening \citep{harding06a}. Doppler broadening gives a Full Width
Half Maximum of \citep{truemper77a,meszaros85a}
\begin{equation}\label{eq:FWHM}
\Gamma_\text{FWHM} =\sqrt{\frac{8\ln(2)kT_\text{e}}{m_\text{e} c^2}} |\cos\theta| \,E_\text{cyc}\,.
\end{equation}
For increasing $\cos\theta$ the line shapes become more asymmetric.
In our simulations the plasma temperature is a free parameter. From
theoretical \citep{lamb90a, isenberg98b} and observational studies \citep{heindl04a}, a relation between observed
magnetic field and temperature has been proposed as
\begin{equation}
  kT_\text{e} \gtrsim 0.2\, E_\text{cyc}
\end{equation}
This would correspond to a temperature for the depicted
setting ($E_\text{cyc} \sim 30\,$\ensuremath{\text{keV}}) of at least $6\,$\ensuremath{\text{keV}}, i.e.\ somewhere in between
the bottom spectrum and the second spectrum from the bottom in Fig.~(\ref{fig:fullspectra_cy_Te}).

\subsection{Continuum shape and photon spawning}
\label{subsect:cont}

\begin{figure}  
  \resizebox{\hsize}{!}{\includegraphics{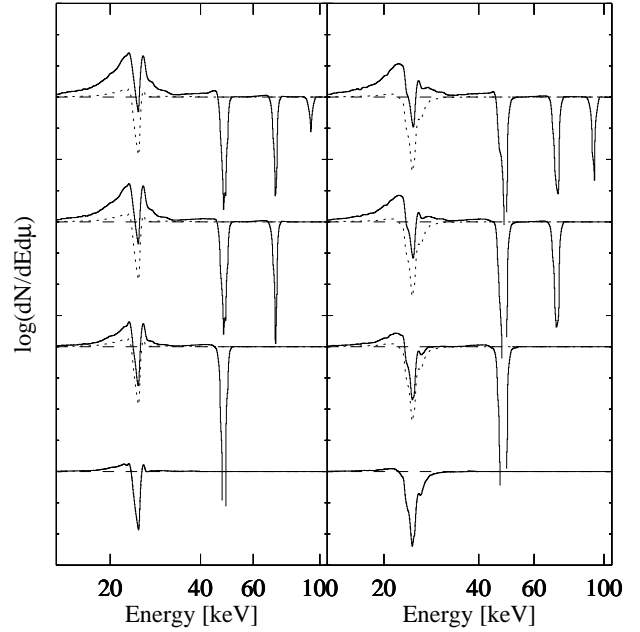}}
  \caption{Changes of the line profiles for consideration of one to three
    harmonics for the scattering processes (bottom to top). The bottom
    graph is overplotted in all cases as a dotted line to guide the
    eye. The input continuum
    spectrum (dashed line) is flat and therefore overemphasizes the effects of
    emission wings. Left panel: cylinder geometry; right panel: slab
    1-1 geometry. In both panels, $kT_\text{e}=3\,$\ensuremath{\text{keV}}, $\cos\theta \in
    [0.125,0.250)$, $\tau_\text{T}=3\cdot10^{-3}$ and $B/B_\text{crit}=0.05$.}
   \label{fig:fullspectra_spawn}
\end{figure}

\begin{figure}
  \resizebox{\hsize}{!}{\includegraphics{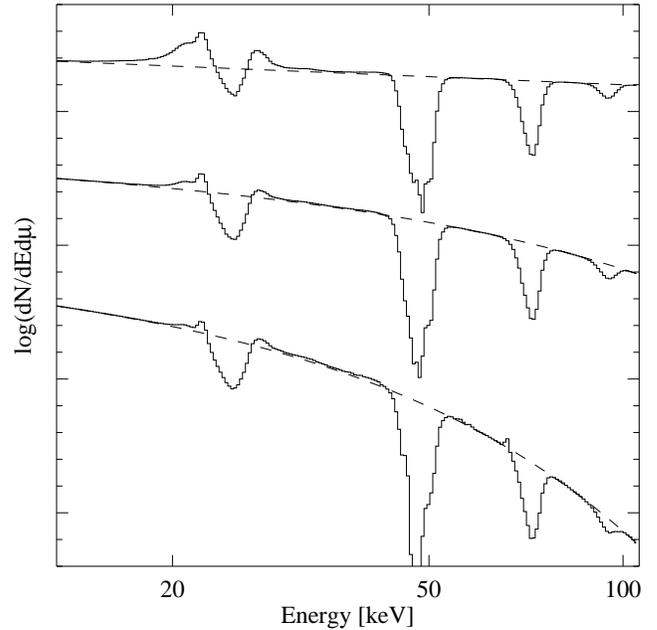}}
  \caption{Line profiles for different continua. All continua have the
    general form of a power law with a photon index $\alpha = 1.0$
    (top). In the middle and bottom plots, the spectra have an
    additional high energy cutoff at the folding energies
    $E_\text{fold}=15\,$\ensuremath{\text{keV}} for the middle and
    $E_\text{fold}=5\,$\ensuremath{\text{keV}} for the bottom plot. We
    show the results for cylinder geometry,
    $kT_\text{e}=3\,$\ensuremath{\text{keV}}, $\cos\theta \in
    [0.375,0.500)$, $\tau_\text{T}=3\cdot10^{-3}$, and $B/B_\text{crit}=0.05$.}
   \label{fig:fullspectra_cy_cont}
\end{figure}

The shape of the cyclotron lines is sensitive to the continuum shape.
In particular, the fundamental line shape and its emission features
vary significantly. For better illustration we show the case of internally
  irradiated plasmas where the emission wings are strongest. The dependence of the line shapes on the continuum
can be understood when considering the photon redistribution in
energy, especially due to photon spawning. Figure
\ref{fig:fullspectra_spawn} shows the change of the line profiles for
a flat input continuum spectrum, when allowing only for electron
transitions between the ground Landau state to the first Landau level, or for
photon-electron scattering leading to up to three
harmonics. In the former case, a single absorption line forms. The
more harmonic scatterings are allowed for, the more lines form, while
the fundamental and lower harmonics become shallower with growing
emission wings. Integrating the photon flux only over the
energy range including just the fundamental line and its emission
wings ($E \le 18\cdot \,\text{keV}\, B_{12}$), we find that the spawned photons
account for as much as $34/64/73$ \% ($n \le 2,3,4$) of the flux for
cylinder and for $11/32/43$ \% of the flux for slab 1-1 geometry.
Considering the whole energy range, the percentage of spawned photons
is $25/52/65$ \% of the total flux for cylinder and $3/15/25$ \% of
the total flux for slab geometry. Note that these numbers are
representative of the extreme and fairly unrealistic case of a flat
input continuum. However, they illustrate well that the line shapes
should change with the spectral hardness of the incident continuum,
where harder spectra exhibit more emission features near shallower
lines. Line profiles for a power law with a photon index $\alpha=1$ with
exponential rolloff at different folding energies
$E_\text{fold}=5,15\,$keV, and for a pure power law continuum spectrum are shown in Fig.\ 
\ref{fig:fullspectra_cy_cont}. In this case, the number of photons
around the first line increases by a factor of $1.1$ and $2.2$ for
the two harder spectra with respect to the softer one.

\section{Applications}\label{sect:applications}
\subsection{A new \textsl{XSPEC} model for cyclotron lines: \texttt{cyclomc}}
At present, a physical model for the analysis of cyclotron lines is
lacking. Typically, for each line independently, multiplicative
phenomenological model components such as Gaussian
, or 
Lorentzian absorption lines are used. They yield simplified,
smoothed line shapes; various CRSFs are fitted independently from each
other one by one. Although it is possible to fit observed data well
with this phenomenological approach, the loss of physical information
is very unsatisfactory. Overlaying several Gaussians or
Lorentzians renders it also difficult to infer the magnetic field
strength with the line centroid position, and does not permit one to
distinguish, e.g., thermal and other broadening of the CRSFs.  The
fact that the line positions are independent of each other in this
approach invokes further information loss. With the increasing quality
of the energy resolution of todays' detectors it is possible to see
more details of the line features calling for a physical model
instead of a phenomenological one.

For purposes of data analysis, we have implemented a cyclotron line
convolution model called \texttt{cyclomc} as a local model for fitting CRSFs in \textsl{XSPEC}. 
Based on the Green's functions of
Sect.~\ref{subsect:technical}, \texttt{cyclomc} is designed to
fit up to four CRSFs simultaneously. Not only the magnetic field
strength and the temperature, but also the optical depth and the ratios of
line positions, line widths and line depths are determined by the
underlying physical picture. The quality of the fit hence permits
conclusions on the accuracy of this picture. As shown
in Sect.~\ref{subsect:res:B}, the line ratios could be a sensible
indicator for instance for magnetic field variations along the line of
photon propagation. The model is
obviously restricted to the physical assumptions and the parameter input chosen
for the Monte Carlo simulations. 
During fitting, it interpolates the results on the assumed input
parameter grid which has been described in
Sect.~\ref{subsect:technical}. Since \texttt{cyclomc} is a convolution model,
self-consistent line shapes are obtained for any given continuum.

\subsection{Observability of the line features}\label{subsect:observability}
\begin{figure}
  \resizebox{\hsize}{!}{\includegraphics[angle=0]{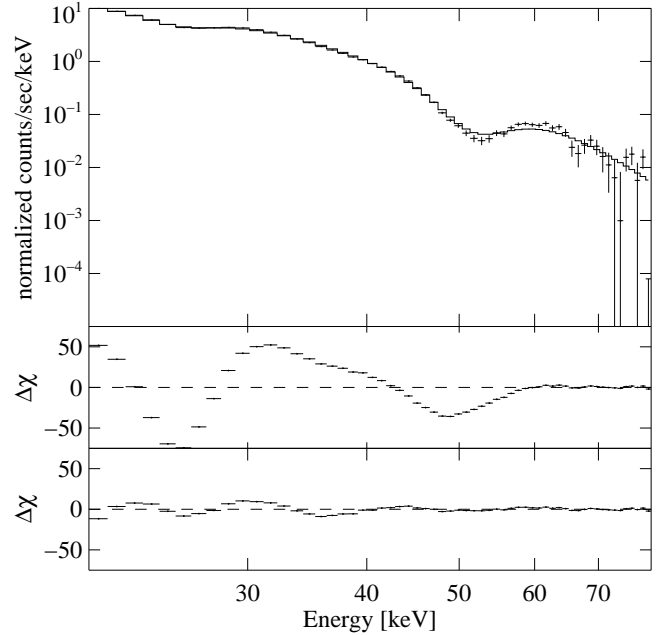}}
  \caption{\textsl{HEXTE} simulations. Upper panel: A simulated spectrum of the
    form \texttt{npex} $\cdot$ \texttt{cyclomc}
    (slab 1-1 geometry, $B_{12}=3.0$, $kT_\text{e}=5.5\,$keV,
$\tau_\text{T}=5\cdot 10^{-4}$, $\cos\theta=0.5$, $z=0.3$, $\alpha_1=1.5$,
    $\alpha_2=1.5$, $E_\text{fold}=6.16\,$keV,
    $A=4.6\cdot 10^{-3}$, $f=1.17$) is fitted with a \texttt{npex}
    multiplied with two Gaussian absorption lines (\textsl{XSPEC}: \texttt{gauabs}). Upper panel: data and fitted
    model. Middle panel: residuals for the simulated spectrum fitted
    with a \texttt{npex} continuum. Bottom panel: residuals for the continuum
    and lines fit shown in the upper
    panel. The strong emission wings of the fundamental line in the
    simulated spectrum are clearly observable in the residuals.}
   \label{fig:simfit_hexte}
\end{figure}

Before applying \texttt{cyclomc} to an example of observational data, we want to
address the more general question of the observability of the
predicted CRSFs in real source data. It is a fundamental question to
assess which features of the model still appear when folded with the
detector responses of todays' observatories. This question had already
been posed by \citet{isenberg98b} considering the issue of different
geometries and densities related to the prediction of significant emission
wings. However, we are not aware of a related study up to date. In
Fig.~\ref{fig:simfit_hexte}, we show the simulated spectrum for a source
which we assign an ideal spectrum of the form of the \texttt{npex}
model \citep{mihara95a,makishima99a}, a negative and positive power law with a
common high-energy cutoff,
\begin{equation}
F(E)=A (E^{-\alpha_1} + f\cdot E^{+\alpha_2}) \exp\left(-\frac{E}{E_\text{fold}}\right)\,,
\end{equation}
folded with a chosen set of the Green's
functions \texttt{cyclomc} for slab 1-1 geometry. The flux is appropriate
for typical HMXB observations. For an assumed observation time of $20\,$ks this
spectrum was folded with the \textsl{RXTE} \textsl{HEXTE}
response and background was added. 
We then fitted our fake
spectrum, using a \texttt{npex} component for the continuum and
two Gaussian absorption lines \citep{coburn02a} to model the simulated
CRSFs. Fig.~\ref{fig:simfit_hexte} shows this fit as well as the residuals for a fit of the continuum component
and for a fit of continuum and line components. The
residuals from the continuum shape represent our simulated line shapes
as we have used the same continuum component for the simulated and for
the model
spectrum. Fitting the first two lines with Gaussians, the emission
wings stay very pronounced in the residuals. For our simulated
spectrum we expect to observe the fundamental CRSF at
$E_1^\text{obs}=26.1\,$\ensuremath{\text{keV}} ($B_{12}=3.0$, $z=0.3$,
$\cos\theta=0.5$). The Gaussian fit of the fundamental line
gives a centroid energy of $24.77\pm 0.03\,$\ensuremath{\text{keV}}
instead. This results indicates that the asymmetric line shapes could
introduce a systematic uncertainty in line energy when modeling
observed data with Gaussian or Lorentzian shapes. We also note that the lines
seem very prominent even for low values of continuum optical depth.

\begin{figure}
  \resizebox{\hsize}{!}{\includegraphics[angle=0]{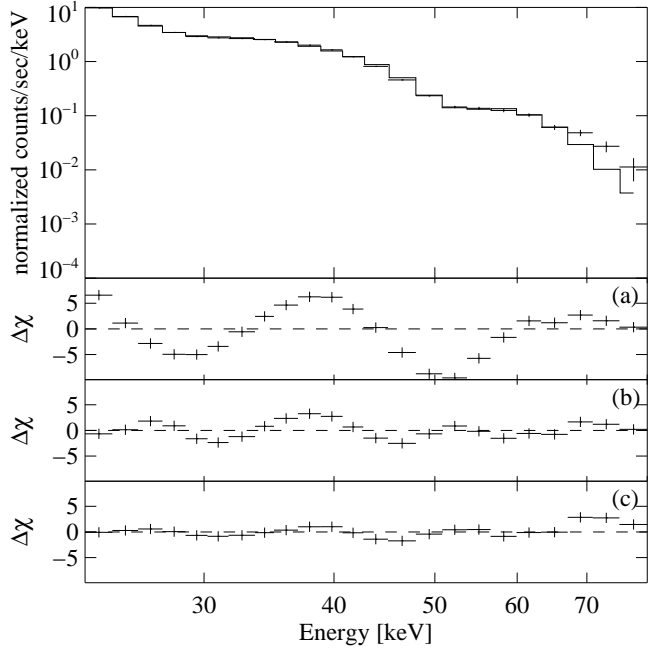}}
  \caption{Fit of the Green's functions CRSF model
  \texttt{cyclomc} to the line features of a
  phase-averaged spectrum of V0332$+$53 obtained from \textsl{INTEGRAL} IBIS observations. The
  data and fitted model (upper panel) along with $\Delta\chi$ residuals are
  shown (three lower panels). The residuals are depicted for
  (a) a continuum fit, and for two different continuum and line fits using (b) slab 1-1
  geometry,
  and (c) slab 1-0 geometry combined with partial covering. The continuum shape is taken as a power law with a
  Fermi-Dirac cutoff (\texttt{fdcut}). For the best fit parameters, see
  Table~\ref{tab:fitpar}.}
   \label{fig:linefit_inte}
\end{figure}

\subsection{Comparison to observational data}\label{subsect:comparison}

During the recent outbursts of the
transient sources V0332$+$53 and A0535$+$26
  \citep{pottschmidt05a,kreykenbohm05a,kretschmar05a,terada06a} more
  data from accreting X-ray pulsar systems containing strong cyclotron resonance scattering features were
  collected. 
We have fitted \textsl{RXTE} and \textsl{INTEGRAL} data of the January 2005 outburst of
  V0332$+$53 as a first example for the comparison of \texttt{cyclomc} with real source
  data. Discovered by \citet{tanaka83a}, V0332$+$53 was the fourth
  accreting X-ray pulsar system where CRSFs were found
  \citep{makishima90a,makishima90b}. Its third CRSF was discovered in \textsl{RXTE} 
  data \citep{coburn05a,pottschmidt05a} and confirmed by subsequent
  \textsl{INTEGRAL} observations \citep{kreykenbohm05a}.
In Fig.~\ref{fig:linefit_inte} we show a fit of the
  first two CRSFs in the \textsl{INTEGRAL} data of V0332$+$53. The fit of the \textsl{RXTE} data of V0332$+$53
  during the same outburst is not shown as the principal results
  turned out very similar for both instruments. We modeled the data with a power law with a Fermi-Dirac
  cutoff (\texttt{fdcut}),
\begin{equation}
  F(E)= A\cdot E^{-\alpha} \left[ \exp \left(
 \frac{E_\text{cut}-E}{E_\text{fold}} \right) +1 \right]^{-1}\,,
\end{equation} 
 folded with the CRSF Green's functions (\texttt{cyclomc}) for
  different geometries, and assuming 5\% systematics for the IBIS
  data. The best fit was obtained for the slab
  1-0 geometry in combination with the scenario of partial
  covering and gave a reduced
  $\chi_\text{red}^2$ of $2.2$ for 13 degrees
  of freedom. Fig.~\ref{fig:linefit_inte} shows the best-fit model
  along with the data and the residuals for fitting the continuum
  component, fitting \texttt{cyclomc} for slab 1-1 geometry, and
  fitting \texttt{cyclomc} for slab 1-0 geometry, combined with
  partial covering, to the data. The best-fit parameters are listed in Table
  \ref{tab:fitpar}. While the slab 1-0 geometry is set apart from the other
  geometries by exhibiting the weakest emission wings, the scenario of partial covering effectively reduces
  the line depths of the model spectrum (compare
  Sect.~\ref{subsect:observability}). We stress that these fits are preliminary, serving at this stage of the model design as a proof of concept
only and as an important first step to help us in understanding the
physics of CRSF formation. The rather high $\chi^2$ should be seen in the light that
this is the first time that a realistic CRSF model has been fit to real
observational data. The fact that we can assess the physics via a
  direct comparison to data enables us to draw conclusions
  concerning the underlying physical picture. First of all, the use of the 1-1 geometry results in
emission wings which are clearly too pronounced for the
  observed data. We have shown in Sect.~\ref{subsect:observability} that a
  simulated source
  with a spectrum identical to the theoretical one would
  indeed still exhibit detectable emission wings in real data. These
  strong emission features are clearly not present in the spectrum of
  V0332$+$53. We therefore conclude that the 1-1 geometry of an internally
  irradiated plasma is not a completely valid
  physical assumption. Considering the source photons to be injected at the
  bottom of the slab (1-0 geometry) or according to a probably biased distribution of incident photon production
  in the plasma seems to be more realistic. Second, we see as a general problem with the fit that the line depths
\texttt{cyclomc} yields for rather low $\tau_\text{T}$ are
very deep compared to the observations. The scenario of partial
  covering of the emergent radiation can solve this issue. In this case,
only part of the emergent radiation is assumed to pass the region of
line formation and is reprocessed with \texttt{cyclomc}. A different possibility to
  ensure shallower line depths, however, would be the introduction of a magnetic
  field gradient along the $B$-field vector in a cylindrical line-forming region (O.~Nishimura, private
  communication). Besides yielding shallower lines, such a scenario
  might also improve the fit due to variations in the line ratios.

\begin{table}
\caption[Best fit parameters for V0332$+$53, slab 1-0 geometry,
  partial covering.]{Best fit parameters for the data and model shown
  in Fig.~\ref{fig:linefit_inte}. Uncertainties are at the 90\%
  confidence level for one interesting parameter; $\chi^2=28.4$ for 13
  d.o.f.} 
\label{tab:fitpar} \centering
\begin{tabular}{l l l }\hline\hline
name [unit] & fit value \\ \hline
$B_{12}\,\,[B/(10^{12}\,G)]$ & $3.05^{+0.05}_{-0.03}$ \\ 
$kT_\text{e}\,\,$[keV] & $10.2^{+0.3}_{-0.3}$\\
$\tau_\text{T}\,\,[10^{-3}]$ & $3.0^{+0}_{-0.4}$  \\
$\cos\theta$ & $0.06^{+0.02}_{-0}$ \\
$z$ & $0.23^{+0.02}_{-0.02}$ \\
$\alpha$ & $0.94^{+0.08}_{-0.04}$ \\
$E_\text{cut}\,\,$[keV] & $12.8^{+1.4}_{-3.3}$ \\
$E_\text{fold}\,\,$[keV] & $7.5^{+0.1}_{-0.1}$ \\
$A_1$ & $2.1^{+8.7}_{-3.8}$ \\
$A_2$ & $0.87^{+0.04}_{-0.05}$ \\
\hline 
\end{tabular}
\end{table}

\section{Summary and Discussion}\label{sect:summary}

We have performed an in-depth study of the formation of cyclotron
resonance scattering features in the spectra of highly magnetized
accreting neutron stars. In particular, we have discussed the
influence of the magnetic field strength, the plasma temperature, the
angle of radiation and the seed continuum spectral form onto the
shapes of cyclotron line features.

Our study, being based on the Monte Carlo simulation code of
\citet{araya96a,araya99a} and \citet{araya00a}, emphasizes several issues which have been
pointed out by those authors before. Examples are Figs.\
\ref{fig:fullspectra_cy_tau} and \ref{fig:fullspectra_sl_tau}, where
we have discussed the optical depth progression matching our
illustrations to Figs.~4--6 from \citet{araya99a} or
Fig.~\ref{fig:ang_redist} for the study of angular
redistribution similar to the illustrations used in \citet{araya00a}. 
With respect to their previous code our
model is different in the following points: most importantly, we did
not restrict ourselves to the study of hard continua, but chose a
Green's functions approach, thus gaining independence from a priori
chosen forms of incident radiation. We relaxed the geometrical constraints on the line-forming region to include the
case of a bottom-illuminated slab as it has been studied by, e.g., \citet{isenberg98b} and \citet{nishimura05a}. Moreover, the higher resolution in angle and energy binning permits to illuminate the complex form of the fundamental feature in detail. The
time-consuming calculations on a huge parameter grid in principal allow for a systematic
  comparison of our results to observational data, which is
  facilitated by the implementation of our local model \texttt{cyclomc} for XSPEC.

Beyond this, we
were able to confirm several results from other authors obtained
with different numerical approaches.
In Fig.~\ref{fig:ang_redist} (Sec.~\ref{subsect:theory}) we have shown the angular
redistribution of photons for cylinder and slab geometry and for two
different values of the plasma optical depth. The percentage of
redistributed photons per ($\cos\theta$)-bin increases with $\theta$ for cylinder
and decreases with $\theta$ for slab geometry as shown in a similar study of
an internally irradiated line forming region by
\citet{araya00a}. For a higher
optical depth and for cylinder geometry, however, the photon
distribution flattens, similar to results shown by \citet{isenberg98b} for
the comparable case of $\Phi = \pi/2$ and slab 1-1 geometry. The
study of the variation of the cyclotron line ratios for a non-uniform magnetic
field picks up an idea from \citet{nishimura05a}, who investigated
this case for a $B$-field which varies linearly with height, slab geometry, and for
similar optical depths. We have confirmed the trend of an increase of
the line ratios with a decreasing magnetic field and a decrease of the line
ratios with an increasing magnetic field within a line-forming
region of slab 1-0 geometry (Fig.~\ref{fig:Bvarying}).

In section \ref{subsect:theory} we reported theoretical predictions of
the model independently from observational data analysis. Except
for the study of the line ratios, outlined above, all analysis assumed a uniform
magnetic field in the line-forming region. A key result is the study of the 
variation of the line parameters of the fundamental feature with
the magnetic field (Fig.~\ref{fig:fullspectra_cy_B}), optical depth and angle
(Figs.~\ref{fig:fullspectra_cy_tau}--\ref{fig:ang_linevary}), and temperature (Fig.~\ref{fig:fullspectra_cy_Te}), which is consistent with
predictions from other authors. We do not see a significant variation
of the fundamental line energy, $E_\text{cyc}$, with the angle, and thus must exclude this simple
scenario as an explanation for observed phase dependent variations of
$E_\text{cyc}$. Omitting the emission wings, the depth of the
fundamental with
respect to the continuum flux is rather stable over $\theta$ whereas
the line width varies significantly with $\theta$ for cylinder and
slab geometry. For both geometries the lines become wider towards
higher $\cos\theta$; for slab geometry an initial decrease for
$\cos\theta < 0.25$ is observed. The variation of the overall line features for different magnetic
field strengths and different temperatures was investigated. Obviously, the positions of the CRSFs are directly linked
to the $B$-field strength (see Eqs.~\ref{eq:12B12} and \ref{eq:Ecycrel}). However, Fig.~\ref{fig:fullspectra_cy_B}
also shows that changes with $B$ as to
the line shapes are rather insignificant. On the other hand, the line
shapes vary strongly with increasing temperature, where more
asymmetric, Doppler-broadened lines arise for higher plasma temperatures.
Furthermore, we have studied in depth variations of the line shapes
with the incident continuum shape. In particular, the shape of the fundamental
line changes with the continuum shape, an effect which can be
understood from photon redistribution, mainly due to photon spawning
in hard continua. As a result, for hard
spectra for instance the emission wings are much more pronounced than
for softer continua. This dependence of the CRSFs on the continuum 
in principle also allows for conclusions on the continuum shape when
modeling cyclotron resonance scattering features.

This study aims at meeting the interests of observers in analyzing
cyclotron lines. We
therefore chose a broad approach to the topic of line formation. Fitting CRSFs in
V0332$+$53 with \texttt{cyclomc} indicates that we can indeed assess CRSFs
in real observed source data with a physical model. This is the first
time such a simultaneous fit of several CRSFs with a realistic,
physical model has been attempted. As outlined in Sect.~\ref{subsect:comparison}, modifications of the
underlying physical scenario guided by these preliminary fits
should help in further improving the comparison of the model and
real data. In Sect.~\ref{subsect:observability} we have assessed the
  general question of the observability of the theoretically predicted line
  shapes. At the early stages of
cyclotron line observations it was not clear whether emission features
-- if they were present in the data -- would be observed or just smeared out by the
detectors. For instance \citet{isenberg98b} and \citet{nishimura05a}
observed that the scenario of a radiation source at the bottom of a
slab as geometry for the line forming region leads to less emission features in the spectra than an
internally irradiated plasma. We have shown in section \ref{subsect:observability} that such strong emission
features as predicted by our scenario should indeed be observable by
the instruments on todays' observatories. The fit results from V0332$+$53 data demonstrate
that large emission wings can be ruled out for this spectrum.

This work is
ongoing, aiming at further generalizations of the CRSF model. We intend to permit angular anisotropy of the continuum photon
flux. \citet{araya00a} have already studied this case for special
angular distributions like peaked emission. We strive for the calculation of
our Green's functions independently not only of the continuum energy but also
of the continuum angular distribution to generalize our model to arbitrary
angular distributions of the incident photons. The first
  issue to be investigated further, however, will be the observed discrepancy in theoretical and
  observational line depths. More realistic $B$-field
  gradients could
  account for shallower lines. We have shown that the consideration of a scenario
 of partial covering, which effectively reduces the
 line depths, significantly improves the fit
 quality. We therefore think that the consideration of non-constant magnetic fields
within the line forming region will
play a major role in the process of better modeling and
understanding cyclotron lines. This issue will be
  subject of a forthcoming paper once having realized a
  systematic comparison of \texttt{cyclomc} for different scenarios for line
  formation to a larger set of observational
  data.

\begin{acknowledgements}
  We thank Rafael Araya for stimulating discussions and for providing
  the first version of the Monte Carlo code used here.  We also thank
  the Department of Physics of the University of Warwick and the
  European Space Astronomy Centre of the European Space Agency for
  their hospitality. We acknowledge the Centre for Scientific
  Computing of the University of Warwick and the Regionales
  Rechenzentrum Erlangen for providing the computing resources used
  in this work. This work was supported by a scholarship from the
  Studienstiftung des Deutschen Volkes and by the DLR grant 50 OR 0302. The observability analysis was based on \textsl{RXTE} data.
  The data comparison was based on observations with \textsl{INTEGRAL}, an ESA project with instruments and science data centre funded by ESA member states (especially the PI countries: Denmark, France, Germany, Italy, Switzerland, Spain), Czech Republic and Poland, and with the participation of Russia and the USA.
\end{acknowledgements}

\end{document}